\documentclass[twocolumn,aps,prd,reprint,superscriptaddress,nofootinbib,letterpaper]{revtex4}

\usepackage{amsmath}
\usepackage{amsfonts}
\usepackage{amssymb}
\usepackage{graphicx, nicefrac}
\usepackage{epsfig}
\usepackage{color}
\usepackage{multirow}
\usepackage{todonotes}
\usepackage{hyperref}
\usepackage[normalem]{ulem}

\usepackage{empheq} 

\def\bea{\begin{eqnarray}}
\def\eea{\end{eqnarray}}
\def\bef{\begin{flalign}}
\def\eef{\end{flalign}}

\def\({\left(}
\def\){\right)}
\def\[{\left[}
\def\]{\right]}
\def\<{\left\langle}
\def\>{\right\rangle}

\newcommand{\weff}{w_{\mbox{\tiny{eff}}}}
\newcommand{\bweff}{\bar{w}_{\mbox{\tiny{eff}}}}

\usepackage{xcolor} 

\begin{document}


\title{Circumventing the Ricci-inverse no-go theorem with\\ complexifiable singularities: a novel dark energy model\\[1ex]
\textnormal{\small{\emph{In memory of Prof.\,A.Vitturi,  for his kind and always helpful soul.}}}}

\author{Mattia Scomparin}
\email{mattia.scompa@gmail.com}
\affiliation{Mogliano Veneto, 31021, Italy}

\date{\today} 

\allowdisplaybreaks

\begin{abstract}
Ricci-inverse gravity is a new type of fourth-order gravity theory based on the anti-curvature tensor, that is, the inverse of the Ricci tensor.
In this context, we introduce a novel method to circumvent the binding effects of a well-known no-go theorem for cosmic trajectories that cannot smoothly join a decelerated cosmic age with the current accelerated expansion of the universe.
We therefore design a new class of Ricci-inverse theories whose cosmologies, without falling into no-go singularities, achieve the observed expansion as a stable attractor solution. This new perspective retrains Ricci-inverse cosmologies as viable dark energy models.
\end{abstract}

\maketitle


\section{Introduction}
\label{section:intro}

Many cosmological observations require a modification to Einstein's theory of General Relativity (GR) to explain the late-time accelerated expansion \cite{SupernovaSearchTeam:1998fmf,Abbott_2019,Pan-STARRS1:2017jku} and the early-time cosmological inflation of our universe \cite{Starobinsky:1980te,PhysRevD.23.347,Hinshaw_2013, Malik:2023yaj}.

There are many ways to construct alternative theories to GR. A possible approach in this direction consists of constructing appropriate modified gravity (MG) theories, in which an alteration of the well-known Einstein Hilbert  (EH) action is considered \cite{Nojiri:2017ncd}. On the other hand, most of these reformulations require troublesome assumptions in their foundations, such as the definition of dimensional coupling constants to be finely-tuned or the introduction of extra dynamical fields that have no obvious relation with  Einstein's strictly geometric formulation of GR. 

Motivated by the above considerations, Amendola \emph{et al.} proposed in Ref. \cite{AMENDOLA2020135923} a  MG theory called Ricci-inverse gravity whose gravitational Lagrangian density $\mathcal{L}_g(R,A)$ is constructed by combining the Ricci scalar $R$ with a new purely geometrical object called \emph{anti-curvature scalar} $A=g_{\mu\nu}A^{\mu\nu}$. By construction, the anti-curvature scalar is the trace of \emph{anti-curvature} tensor $A^{\mu\nu}=R_{\mu\nu}^{-1}$, which is the inverse of the Ricci tensor $R_{\mu\nu}$ defined by the relation
\begin{equation}
\label{eq:antiscalar}
A^{\mu\sigma}R_{\sigma\nu}=\delta^\mu_\nu\,.
\end{equation}
Here, $g_{\mu\nu}$ is the metric tensor and $\delta^\mu_\nu$ is the Kronecker delta function.
For the avoidance of doubt, one must pay attention to the fact that $A\ne R^{-1}$. However, focusing only on the physical dimension, $[A]=[R]^{-1}$.

As a consequence, one can use the anti-curvature scalar to introduce new Lagrangian terms  into EH action which, resulting directly from $R$, share its same physical dimension and do not invoke any new dynamical field to modify GR. Among all scale-free possibilities, we mention: (i) the pure anti-curvature term $A^{-1}$, (ii) the general power-law profile $A^\ell R^{\ell+1}$, and (iii) any general profile of the form $g(RA)R$. Note that (i) belongs to (ii) when $\ell=-1$, while (iii) reduces to (ii) when $g(RA)=(RA)^\ell$.

In recent years, Ricci-inverse gravity has been extensively studied in the currently available literature.
Examples of applications include the investigation of: (i) anisotropic \cite{Shamir:2022mgq}, compact \cite{Shamir:2023ouh} and charged \cite{Malik:2023hhn, Malik:2024zhb} star structures \cite{Malik:2024uti, Malik:2023yaj},  (ii) matter-antimatter asymmetry phenomena  \cite{Jawad:2022eoj}, (iii) novel aspects related to wormhole \cite{Mustafa:2024ark} and blackholes \cite{Ahmed:2024jcb} solutions, (iv) axially symmetric spacetimes with causality violation \cite{deSouza:2023egd, deSouza:2024vqw}, and (v) cosmic structures within a Sub-Horizon non-relativistic Weak-Field limit \cite{Scomparin:2021zim}.
Although most of these works emphasize the increasing relevance and applicability of Ricci-inverse gravity in diverse scenarios, two binding no-go theorems have been shown to rule out many cosmological realizations of this theory.

The \emph{first no-go theorem} concerns the late-time accelerated expansion phase of universe. Its original formulation states that any Lagrangian density $\mathcal{L}_g(R,A)$ containing terms proportional to $A^\ell$, with any positive or negative $\ell$, cannot smoothly join a cosmic decelerated era with the current accelerated  expansion of universe \cite{Planck:2018vyg} (see Refs.\cite{AMENDOLA2020135923, Das:2021onb} for a detailed discussion). 
In fact, assuming a standard Friedmann-Lema\^{i}tre-Robertson-Walker (FLRW) metric, between these two epochs $A$ must pass through both $0$ and $\pm\infty$ values, leading a generic power of $A$ to blow up due to singularities.  
Up to now, attempts to find loopholes to escape this no-go theorem have not proved satisfactory. Indeed, 
it has been shown that the shiftings introduced by spatially curved metrics or anisotropic backgrounds are too small to move the singularities sourced by $A$ outside the observational range. In the end, only well-though non-polynomial Lagrangians (i.e. respecting the property $[\mathcal{P}1]$ of being regular both for $A\rightarrow 0$ and $A \rightarrow \pm \infty$)
seem promising enough to circumvent the problem, but they are very complicated and consequently have not been investigated so far \cite{AMENDOLA2020135923,Das:2021onb}.

Another cosmic scenario now forbidden to Ricci-inverse gravity is the so-called inflationary era, which occurred very rapidly after the Big Bang \cite{Starobinsky:1980te, Guth:1980zm, Linde:1981mu}. Although this primordial phase is not affected by singularities from anti-curvature terms,  a stability analysis based on a dynamical system method showed that, even when $\mathcal{L}_g(R,A)$ depends on simple powers of $A$ (e.g $A$ or $A^2$),  all isotropic inflationary solutions turn out to be unstable with respect to field perturbations \cite{Do:2020vdc,Do:2021fal}.
This behavior is due to the fact that the existence domain and the stability region of the solutions do not overlap. Therefore, the impossibility to have a stable (isotropic) inflationary solutions as well as a smooth exit from inflation era (see, e.g. Ref. \cite{Das:2021onb}) in elementary Ricci-inverse cosmologies is essentially the content of the \emph{second no-go theorem}.

In the absence of effective loopholes, all these results have raised great doubts about the cosmological viability of Ricci-inverse gravity to be a suitable inflationary model or DE candidate. 
To the author's knowledge, no valid proposal has been found at present  to circumvent the action of the two no-go theorems. Furthermore, neither non-trivial extensions involving higher-order scalar combinations in the anti-curvature tensor\footnote{Here, $A_\nu^{\alpha\rho\gamma}$ is the anti-Riemann tensor, defined by the identity  $A^{\alpha\rho\gamma}_\nu R^\mu_{\alpha\rho\beta}=\delta^\mu_\nu\delta^\gamma_\beta$.}   such as  $\mathcal{L}_g(R,A^{\mu\nu}A_{\mu\nu})$ or $\mathcal{L}_g(R,A^{\alpha\beta\gamma}_\mu A_{\alpha\beta\gamma}^\mu)$ seem a promising prospect for treating in a simple way the pathologies that afflict  the Ricci-inverse gravity, which, in the end, is still ruled out \cite{Das:2021onb}.

At this point, one might come to the conclusion that there is no way around no-go theorems for Ricci-inverse cosmologies.
Of course, every no-go theorem is based on arguments that one can try to break or reconsider. For example, in the case of the first no-go theorem, the most suitable point to be reconsidered is that $\mathcal{P}1$ is really the only property that makes non-polynomial Lagrangians capable of realizing regular equations of motion.
In fact, looking for a possible workaround of the theorem in this way, the author realized that there is indeed a new property that  can regularize the no-go singularities. 

Generally speaking, our reasoning is as follows. Focusing on Ricci-inverse cosmologies studied so far, we noticed that their Lagrangians exhibit no-go singularities belonging only to the real axis  $\mathbb{R}$. However, if such a Lagrangians could be redrawn to push the no-go singularities into the complex plane $\mathbb{C}$,
we might of course expect the effects of the first no-go theorem to become inoffensive, since the cosmic trajectories are and remain purely real. We call this property $[\mathcal{P}2]$, i.e. \emph{regularizability with complexifiable singularities}. It refers to the characteristic of some non-polynomial Lagrangians, under certain conditions on their free parameters, to shift their no-go singularities to the complex plane with a non-zero imaginary part.

Driven by this idea, the aim of this paper is twofold: (i) to review the known aspects about the first no-go theorem in Ricci-inverse gravity in the form of $\mathcal{L}_g(R,A)$,  and (ii) to present new original results on this topic. 
 In particular, using $[\mathcal{P}2]$, we present a novel singularity-free DE model whose cosmic trajectories smoothly connect a cosmic decelerated era with the current accelerated expansion of the universe, which in turn  emerges as a stable attractor solution alongside a de Sitter phase.

The outline of our work is as follows. 
In Section \ref{sec:gravTH} we begin by briefly introducing the full Ricci-inverse theory $\mathcal{L}_g(R,A)$ and discussing the related covariant field equations. 
In Section \ref{sec:gravTHnogo} we introduce the Ricci and anti-curvature scalars in a flat-space FLRW universe and, contextually, in Subsection \ref{sec:nogoRev}  we review the state of the art on the first no-go theorem. 
Section \ref{sec:class} is devoted to the introduction of a novel class of Ricci-inverse models that might be expected to violate the first no-go theorem. Such models depend on two free-parameters and in Subsection \ref{sec:circumv} we explain how to implement our property $[\mathcal{P}2]$ to tune them and obtain healthy Ricci-inverse cosmological solutions. 
After finding the related modified Friedmann equations in Section \ref{sec:cosmoEq}, in Subsection \ref{sec:Plaw} and Subsection \ref{sec:presslessM} we provide proofs of stable de Sitter and power-law solutions (compatible with the current accelerated phase).
Finally, in Section \ref{sec:concl} we summarise our results and identify future avenues of investigation..

In this work, we use the metric signature $(-,+,+,+)$ and Einstein's convention on repeated dummy indices is assumed.


\section{Ricci-inverse gravitational theory}
\label{sec:gravTH}

Let us consider the full Ricci-inverse theory described by the basic action
\begin{equation} 
\label{eq:lag}
S=\int\!\left(\frac{1}{2\kappa}\mathcal{L}_g+\mathcal{L}_m\right)\sqrt{-g}\,dtd^3\!x\,,
\end{equation}
where $g$ is the determinant of the metric $g_{\mu\nu}$ and $\kappa\equiv8\pi G/c^4$ is a constant that depends on the gravitational parameter $G$ and the speed of light $c$. The first term $\mathcal{L}_g=\mathcal{L}_g(R,A)$ corresponds to the MG Lagrangian density, which is an \emph{a priori} arbitrary function of the Ricci and anti-curvature scalars. The second term $\mathcal{L}_m$ is the matter Lagrangian density, that we assume minimally  coupled  with the metric tensor only. 

It is immediate to check that if $\mathcal{L}_g= R$, then Eq. \eqref{eq:lag} reduces to the usual EH action for GR.

Varying the above action with respect to the metric, the field equation is $\delta S=0$  \cite{AMENDOLA2020135923, Lee_2021}. By differentiating definition \eqref{eq:antiscalar} and using integration by parts we obtain
\begin{equation}
\label{eq:varS}
0=\frac{\delta S}{\delta g^{\mu\nu}}\delta g_{\mu\nu} =\frac{1}{2} \int\!\sqrt{-g}\,\big[\mathcal{G}^{\mu\nu}-\kappa T^{\mu\nu}\big]\delta g_{\mu\nu}\,dtd^3\!x\,,
\end{equation}
where we introduced the modified Einstein tensor 
\begin{align}
\mathcal{G}^{\mu\nu}&\equiv\partial_R\mathcal{L}_g R^{\mu\nu}
- \tfrac{1}{2} \mathcal{L}_g g^{\mu\nu}\nonumber\\
&- \partial_A \mathcal{L}_g A^{\mu\nu}\nonumber\\
&- \tfrac{1}{2} \nabla^{\alpha}{\nabla_{\alpha}{(\partial_A \mathcal{L}_g A_{\sigma}^{\mu}A^{\nu\sigma}})}
\\
&+g^{\rho\mu} (\nabla_{\alpha}{\nabla_{\rho}{\,\partial_A \mathcal{L}_g}}) A^{\alpha}_{\sigma} A^{\nu\sigma}
\nonumber\\
&- \tfrac{1}{2} g^{\mu\nu} \nabla_{\alpha}{\nabla_{\beta}{(\partial_A \mathcal{L}_g A_{\sigma}^{\alpha} A^{\beta\sigma})}}
\nonumber\\
&- \nabla^{\mu}{\nabla^{\nu}{\partial_R\mathcal{L}_g}}
+ g^{\mu\nu}\nabla^{\alpha}{\nabla_{\alpha}{\,\partial_R\mathcal{L}_g}}  \,,\nonumber
\end{align}
and the matter  energy-momentum tensor 
\begin{equation}
\label{eq:Tm}
T^{\mu\nu}\equiv\frac{2}{\sqrt{-g}}\frac{\delta(\sqrt{-g}\mathcal{L}_m)}{\delta g_{\mu\nu}}\,.
\end{equation}

Since equation \eqref{eq:varS} must apply for any variation $\delta g_{\mu\nu}$, this implies that the tensor field equation becomes
\begin{equation}
\label{eq:ModG}
\mathcal{G}^{\mu\nu}=\kappa\, T^{\mu\nu}\,.
\end{equation}

In our notation, the $\nabla_\mu$ symbol is understood as the covariant derivative. Additionally, unless otherwise specified, we use the symbols $\partial_R\equiv\partial/\partial R$ and $\partial_A\equiv\partial/\partial A$ to denote  derivative with respect to $R$ or $A$, respectively. 
 
We recall that many functional forms for $\mathcal{L}_g$ have been proposed to obtain cosmic solutions within the Ricci-inverse theory \cite{AMENDOLA2020135923,Das:2021onb}. 

However, as we are going to show in the next Section \ref{sec:gravTHnogo}, such models generally fail to explain the late-time acceleration phase of the Universe since their trajectories encounter no-go  singularities.


\section{Ricci-inverse FLRW cosmologies}
\label{sec:gravTHnogo}

On sufficiently large scales, the Cosmological Principle states that the universe is uniformly isotropic and homogeneous. The most general space-time compatible with such assumptions is parametrized in terms of the FLRW line-element, which in polar coordinates is given by
 \begin{equation}
\label{eq:g_FLRW}
ds^2=-c^2dt^2+a^2\!\left(dr^2+r^2d\Omega_2^2\right)\,,
\end{equation}
where $a(t)$ is the scale factor and $d\Omega_2^2=d\theta^2+\sin^2\!\theta\, d\varphi^2$ is the tridimensional solid-angle element. 

We point out that in \eqref{eq:g_FLRW} we assume the flat-space case. In fact, being our study  focused on DE (and inflationary) models, observations show that in these casuistics the space-curvature density parameter $\Omega_k$ is very close to zero \cite{Planck:2018vyg,Hergt:2022fxk}. Therefore the space-curvature  interferes little or nothing on the actual realization of the no-go theorems \cite{AMENDOLA2020135923,Do:2021fal}.

Now, the expression for the Ricci scalar calculated from the metric \eqref{eq:g_FLRW} is given by
\begin{equation}
\label{eq:R}
R=\frac{6 H^2}{c^2}(\xi+2)\,,
\end{equation}
where the Hubble parameter $H\equiv \dot{a}/a$ quantifies the rate of time-evolution of the scale factor, and the overdot denotes derivative with respect to cosmic time $t$.

For convenience, we have also introduced the new variable $\xi\equiv H'/H= \ddot{a}\,a/\dot{a}^2-1$, where the prime symbol stands for $' = d/d \ln a$.

According to these results, using definition \eqref{eq:antiscalar}, the corresponding anti-curvature scalar has expression
\begin{equation}
\label{eq:A}
A=\frac{2c^2}{3H^2}\frac{6+5\xi}{(1+\xi)(3+\xi)}\,.
\end{equation}

We end up this section by pointing out: (i) that our results \eqref{eq:R} and \eqref{eq:A} confirm once again that $A\ne R^{-1}$, and (ii) that in the Minkowski limit $H\! \rightarrow \!0$ the anti-curvature scalar $A$ becomes singular while $R$ and $A^{-1}$ do not.


\subsection{The first no-go theorem}
\label{sec:nogoRev}

Before proceeding, it is worth briefly discussing a pathology that afflicts FLRW Ricci-inverse cosmologies.

According to expressions \eqref{eq:R} and \eqref{eq:A}, we see that both the Ricci and anti-curvature scalars exhibit some critical points.
In fact, except for special initial conditions such that $H$ vanishes or diverges: (i) $A$ is singular for $\xi_{\mbox{\tiny c,1}}=-1$ and $\xi_{\mbox{\tiny c,3}}=-3$, (ii) $R$ vanishes for $\xi_{\mbox{\tiny c,2}}=-2$, and (iii) $A$ vanishes for $\xi_{\mbox{\tiny c,4}}=-6/5$.
Due to this fact, it is clear that any Lagrangian density $\mathcal{L}_g(R,A)$ containing terms proportional to $A^\ell$ or $R^{-|\ell|}$ (with any positive or negative $\ell$) cannot in general join without problems the cosmic epochs passing through $\xi_{\mbox{\tiny c,i}}$ ($i=1,2,3,4$). Indeed, these terms blow up due to the singularities  sourced by $\xi_{\mbox{\tiny c,i}}$, and this fact has repercussions on the equations of motion and their solutions, which will show singularities at the same cosmic epochs \cite{AMENDOLA2020135923}. 
As we are going to discuss, this fact emerges to be particularly detrimental during the late-time accelerated expansion of universe. 

It can be proven that the $\xi$ variabile can be rewritten 
\begin{equation}
\label{eq:weff}
\xi=-\tfrac{3}{2}\big(1+\weff\big)\,,
\end{equation}
where $\weff$ is the effective Equation of State (EoS) parameter of the universe \cite{AMENDOLA2020135923}.
Recent observations confirm that our universe has evolved from a
decelerated phase $\weff=0$ ($\xi\approx-1.5$) to the current accelerated phase $\weff=-0.685$ ($\xi\approx-0.472$) \cite{PanSTARRS1:2017jku,AMENDOLA2020135923}. Thus, being $\xi_{\mbox{\tiny c,1}},\xi_{\mbox{\tiny c,4}}\in[-1.5;-0.472]$, it is clear that any power of $A$ evolves a singularity between these two epochs. Due to the above restrictions, no power of $R$ can cure such singularities  and this fact results in a no-go theorem rejecting the Ricci-inverse gravity to be a suitable DE candidate to smoothly joint these two epochs.

\begin{figure}[t]
\includegraphics[width=9.5cm]{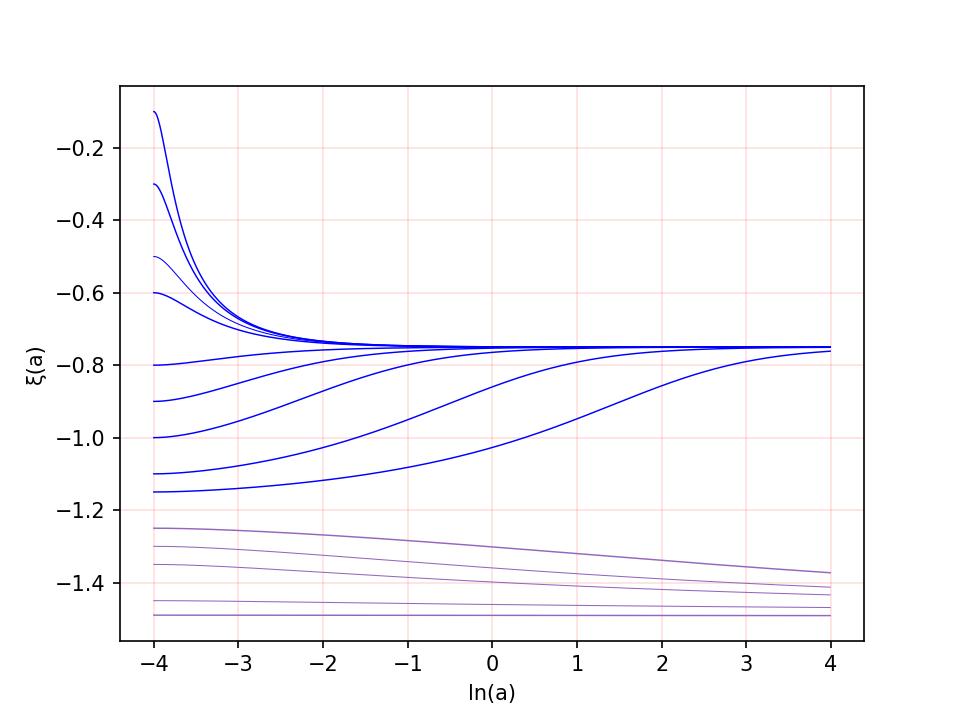}
\centering
\caption{Numerical solutions $\xi(a)$ for Eq. \eqref{eq:summ2} in the pressureless matter case. Here, $\alpha=-4$ and $\beta=0$ do not satisfy the singularity-free condition \eqref{eq:singfree}. It is evident that the solutions $\bar{\xi}=-1.5$ (magenta lines) and $\bar{\xi}=-0.75$ (blue lines)  are stable attractor. However, as discussed in Section \ref{sec:nogoRev}, a divide at $\bar{\xi}=-1.2$ realizes the effects of the no-go theorem.}
\label{img:nogo}
\end{figure}

Evidence of this fact is well illustrated in Fig. \ref{img:nogo}, where we plot the cosmic solutions for the Ricci-inverse theory $\mathcal{L}_g=R-4A^{-1}$. Here, the expected singularity  sourced by the $A^{-1}$ term is $\xi_{\mbox{\tiny c,4}}=-1.2$, which clearly appears as a divide of the solutions. Consequently, due to the $\xi_{\mbox{\tiny c,4}}$ ridge, the cosmic evolution cannot smoothly pass from $\xi\approx-1.5$ to $\xi\approx-0.472$.

To author's knowledge, no valid proposal has been currently found to simply circumvent the action of the first no-go theorem  \cite{AMENDOLA2020135923,Das:2021onb}. However, in searching for a possible solution to the problem, the author realized the existence of a new property that could regularize no-go singularities.


\section{A novel dark energy model with complexifiable singularities}
\label{sec:class}

In this Section we introduce the notion of \emph{regularizability with complexifiable singularities} $[\mathcal{P}2]$ and show explicitly how it allows us to get rid of the no-singularities.
In mathematical terms, this property  refers to the characteristic of certain non-polynomial functions to be able to shift, under certain conditions on their free parameters, their (real) singularities from the real axis $\mathbb{R}$ to the complex plane $\mathbb{C}$.
\begin{figure}[t]
\includegraphics[width=7cm]{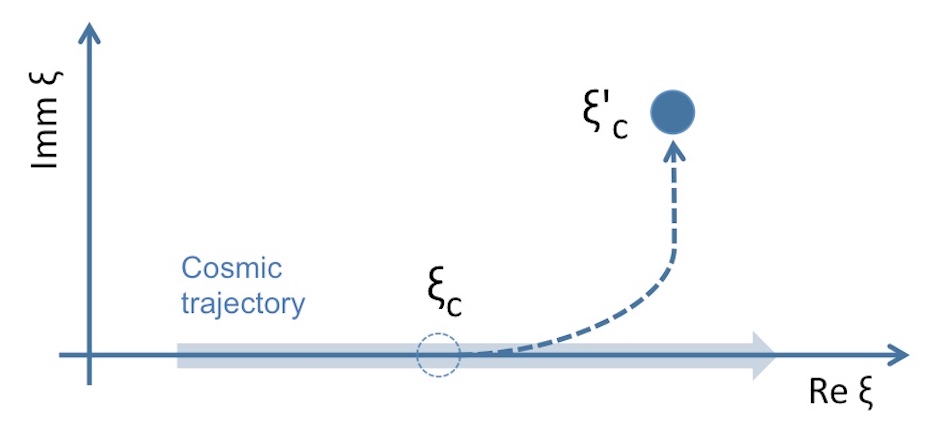}
\centering
\caption{Schematic representation of the circumvention method of the first no-go theorem using complexifiable singularities. If $\mbox{Imm}\,\xi'_{\mbox{\tiny c,i}}\ne0$, then the first no-go theorem becomes inoffensive since the cosmic trajectories are real.}
\label{img:complexify}
\end{figure}

With this idea in mind, we can argue as follows.
In the case of Ricci-inverse gravity we observe that $\xi_{\mbox{\tiny c,i}}\in \mathbb{R}$, $\forall\,i=1,2,3,4$. Hence, if we able to designing a Lagrangian density $\mathcal{L}_g$ that satisfies $[\mathcal{P}2]$, then we can transform any troublesome singularity $\xi_{\mbox{\tiny c,i}}\in\mathbb{R}\rightarrow\xi'_{\mbox{\tiny c,i}}=\mbox{Re}\xi'_{\mbox{\tiny c,i}}+j\mbox{Imm}\xi'_{\mbox{\tiny c,i}}\in\mathbb{C}$.
As a consequence, if $\mbox{Imm}\xi'_{\mbox{\tiny c,i}}\ne0$ (see, e.g. Fig. \ref{img:complexify}), then 
the effects of the first no-go theorem become inoffensive, since cosmic trajectories are physical, i.e. they only evolve along the real axis.
In particular, by virtue of this fact, the undesirable singularities that lie between the decelerated and accelerated phases of our universe can be eliminated, resurrecting the Ricci-inverse theory as a suitable DE alternative.

Essential to this scenario is the observation that the design of Lagrangian densities satisfying the $[\mathcal{P}2]$ property is by no means straightforward.
For example, the classes of (ruled-out) theories $\mathcal{L}_g=R-\alpha A^{-1}$ (see, e.g. Refs. \cite{AMENDOLA2020135923,Das:2021onb})   or $\mathcal{L}_g=R-\alpha A$ (see, e.g. \cite{deSouza:2024vqw}), with  $\alpha \in \mathbb{R}$, do not satisfy $[\mathcal{P}2]$. Indeed, there is no value of $\alpha$ able to influence the position of the $\xi_{\mbox{\tiny c,1}}=-1$ and $\xi_{\mbox{\tiny c,4}}=-6/5$ singularities. This fact is easily visible by substituting \eqref{eq:R} and \eqref{eq:A} in the above Lagrangian densities and then making the common denominator.


\subsection{Circumventing the first no-go theorem}
\label{sec:circumv}

A simple (and novel) model illustrating how our method leads to singularity-free cosmologies considers a Lagrangian density of the form
\begin{equation}
\label{eq:theory}
\mathcal{L}_g=R\left[1+\frac{\alpha}{\beta+AR}\right]\,,
\end{equation}
where $\alpha,\beta\in \mathbb{R}$ are constant parameters. 
It is immediate to check that if $\alpha=\alpha_{GR}=0$, then \eqref{eq:theory} reduces to the usual EH action for GR. Besides, when $\alpha\ne0$ and $\beta=0$ we fall back in the (ruled-out) models within the Class Ia studied in Refs. \cite{AMENDOLA2020135923, Das:2021onb}. 
In addition, we stress that $\alpha$ and $\beta$ are dimensionless.  Therefore, our MG model does not require any new dimensional scales, just as it does not introduce any additional fields to those of GR theory.

The strength of our method lies in the fact that, provided $\mathcal{P}2$ is satisfied, it allows us to very easily transform no-go singularities from real values to complex values, thus rendering them physically ineffective. Now we show that this is exactly what happens. We begin by writing down our Lagrangian \eqref{eq:theory} in a FLRW universe. Recalling \eqref{eq:R} and \eqref{eq:A}, we easily find that
\begin{equation}
\label{eq:der}
\mathcal{L}_g\!=\!\frac{6H^2}{c^2}(\xi\!+\!2)\left[1\!+\!\frac{\alpha(\xi\!+\!1)(\xi\!+\!3)}{(\beta\!+\!20)\xi^2\!+\!4(\beta\!+\!16)(4\xi\!+\!3)}\right].
\end{equation}
The no-go singularities are clearly evident in the denominator of \eqref{eq:der} and solve the algebraic equation of the second degree
\begin{equation}
\label{eq:criteq}
(\beta+20)\xi_c^2+4(\beta+16)(4\xi_c+3)=0\,,
\end{equation}
whose discriminant is
\begin{equation}
\Delta=4(\beta+4)(\beta+16)\,.
\end{equation}

We now demand that the theory is non-pathological, in the sense that equation \eqref{eq:criteq} admits only complex solutions. To meet this demand, we impose $\Delta<0$, which finally results in the \emph{singularity-free condition}
\begin{equation}
\label{eq:singfree}
-16<\beta<-4\,.
\end{equation}

From this result, it can be stated that if $\beta$ satisfies condition \eqref{eq:singfree}, then the FLRW cosmologies emerging form our Lagrangian \eqref{eq:theory} are safe from the effects of no-go singularities. Interestingly, this condition does not depend on the value of the coefficient $\alpha$.

As a final remark, we point out that if, instead of $\beta+AR$, in the denominator of \eqref{eq:theory} we had chosen a higher-degree polynomial like $\sum_p[ \beta_p+ (AR)^q]^{p}$, $p,q\in \mathbb{R}$,
then the analogous of equation \eqref{eq:criteq} would become more difficult to discuss being of degree greater than two. Thus, for the illustrative purpose of our study, we decide to adopt the simplest choice \eqref{eq:theory}.


\section{Cosmological equations}
\label{sec:cosmoEq}

The preceding Subsection \ref{sec:circumv} was dedicated to introduce a novel Lagrangian density \eqref{eq:theory} that in principle, by imposing condition \eqref{eq:singfree}, we expect to regulate the no-go singularities plaguing Ricci-inverse cosmologies.
Now we have knowledge of the $\mathcal{L}_g$ profile, we want to find the evolution of
the scale factor and other quantities to verify that this is exactly what happens.

To achieve this, we start with the energy-momentum tensor  \eqref{eq:Tm}, which we assume to be that of a perfect fluid. According to the literature, it can be written as
\begin{equation}
\label{eq:T_FLRW}
T^{\mu\nu}=\left(\rho+ c^{-2}P\right)u^\mu u^\nu+P g^{\mu\nu}\,,
\end{equation}
where $\rho=\sum_i\rho_{(i)}$ and   $P=\sum_i P_{(i)}$ are the total mass-energy density and the total hydrostatic pressure of a mixture of two or more non-interacting perfect fluids, each labelled with the index $i\in I$. It is well known that the set  $I$ can include a variety of fluids, such as: radiation $(r)$ for photons, pressureless non-relativistic matter $(m)$ for baryonic matter  and dark matter, and the cosmological constant $(\Lambda)$ for non-dynamical dark energy.

The mass-energy densities and the hydrostatic pressures that appear in Eq. \eqref{eq:T_FLRW} are all measured in the fluid's (comoving) rest frame, where the fluid's four-velocity $u^{\mu}$ has expression $u^\mu=\!^\top\!(c,0,0,0)$.
Hence, we can then rewrite the above expression as
\begin{equation}
\label{eq:TTT}
T^\mu_\nu=\mbox{diag}\left(-\rho c^2, P,P,P\right)\,.
\end{equation}

In this context, it is useful recalling that the relationship between energy density and pressure is generally described by means of an Equation of State (EoS).
 Accordingly, henceforth we use assume the following $P(\rho)$ barotropic EoS
 \begin{equation}
 \label{eq:EOS}
 P_{(i)}=c^2w_{(i)} \rho_{(i)} \quad (i\in I)\,,
\end{equation}
which applies for each $i$-component of the fluid.
Here, $w_{(i)}\in \mathbb{R}$ are constant parameters. In particular, $w_{(r)}=1/3$ for a radiation density, $w_{(m)}=0$ for a pressureless non-relativistic matter density, and $w_{(\Lambda)}=-1$ for a cosmological constant density.

Having introduced the necessary background, we can now proceed to derive the modified Friedmann equations.

Using the profiles \eqref{eq:g_FLRW} \eqref{eq:theory} and expressions \eqref {eq:TTT} \eqref{eq:EOS} and then applying them to the field equation \eqref{eq:ModG}, the following cosmological equations for the density parameter $\Omega_{(i)}\equiv \tfrac{8}{3}\pi G\rho_{(i)}H^{-2}$ can be obtained after somewhat tedious algebra
\begin{subequations}\label{eq:summ}
\begin{empheq}[left=\empheqlbrace]{align}
	\label{eq:summ1}
       \,&\sum_{i\in I}\Omega_{(i)}=\psi_0+\psi_1 \xi'\,,\\
	\label{eq:summ2}
      \,&\sum_{i\in I}w_{(i)}\Omega_{(i)}=\varphi_0+\varphi_1\xi'+\varphi_2\xi''+\varphi_3(\xi')^2\,,\\
      \label{eq:summ3}
      \,&\Omega_{(i)}'=\chi_{(i)}\Omega_{(i)}\quad (i\in I)\,.
\end{empheq}
\end{subequations}

Strictly speaking, Eqs. \eqref{eq:summ1} and  \eqref{eq:summ2} are derived directly from the $(t,t)$ and the $(r,r)$ components of the field equation \eqref{eq:ModG}, respectively.
Using these, we then also derived the matter conservation equation \eqref{eq:summ3}, which emerges because the left-hand side of Eq.  \eqref{eq:ModG} is divergenceless, i.e. it satisfies the Bianchi identity, when equipped with our profile \eqref{eq:theory}. However, only two of the three Eqs. \eqref{eq:summ} are functionally independent. 

Interestingly, since Eq. \eqref{eq:summ3} depends at most on $\xi''$ and being $\xi\sim\ddot{a}$, our model (as the whole Ricci-inverse framework) is a type of fourth-order gravity.
\begin{widetext}
Our system \eqref{eq:summ} depends on a set of seven $\xi$-dependent functions, which have the following expression
\begin{align}
\label{eq:functXi}
\psi_0(\xi)&=1+\frac{{\alpha (\xi + 3)^2 (\beta (\xi + 1)^2 + 4 (\xi + 2)^2)}}{{(\beta (\xi + 1) (\xi + 3) + 4 (\xi + 2) (5 \xi + 6))^2}} 
\,,\\
\psi_1(\xi)&=-\frac{(8\alpha(\beta(\xi(\xi(5\xi+18)+27)+18)+36(\xi+2)^3)}{\beta(\xi+1)(\xi+3)+4(\xi+2)(5\xi+6)^3}\,,
\end{align}
and
\begin{align}
\varphi_0(\xi)&=-\frac{{(2\xi + 3)(\alpha(\xi + 3)^2(\beta(\xi + 1)^2 + 4(\xi + 2)^2) + (\beta(\xi + 1)(\xi + 3) + 4(\xi + 2)(5\xi + 6)^2)}}{{3(\beta(\xi + 1)(\xi + 3) + 4(\xi + 2)(5\xi + 6))^2}}
\,,\\
\varphi_1(\xi)&=8\alpha\frac{(\xi+2)(\beta(\xi(\xi(5\xi+18)+27)+18)+36(\xi+2)^3)}{(\beta(\xi+1)(\xi+3)+4(\xi+2)(5\xi+6))^3}
\,,\\
\varphi_2(\xi)&=8\alpha\frac{(\beta(\xi(\xi(5\xi+18)+27)+18)+36(\xi+2)^3)}{3(\beta(\xi+1)(\xi+3)+4(\xi+2)(5\xi+6))^3}
\,,\\
\varphi_3(\xi)&=8\alpha\frac{{(-\beta^2(\xi(\xi(\xi(5\xi+24)+54)+72)+45)-8\beta(\xi(\xi+3)(\xi(17\xi+45)+108)+180)-720(\xi+2)^4}}{{\beta(\xi+1)(\xi+3)+4(\xi+2)(5\xi+6)^4}}
\,,
\end{align}
\end{widetext}
and
\begin{equation}
\chi_{(i)}(\xi)=-\big[3\left(w_{(i)}+1\right)+2\xi\big]\quad (i\in I)\,.
\end{equation}
Finally, by defining the  anti-curvature density parameter $\Omega_{(A)}=1-\psi_0+\psi_1 \xi'$,
then Eq. \eqref{eq:summ1} can be rewritten as
\begin{equation}
\label{eq:FLRW_ttO}
\sum_{i\in I}\Omega_{(i)}+\Omega_{(A)}=1\,.
\end{equation}
Now we are ready to examine how various cosmological scenarios can be realized within our  gravitation model. This topic will be addressed in the next section.
%

\subsection{Power-law \& de Sitter solutions}
\label{sec:Plaw}

In this section we consider scenarios in which the scale factor varies like (i) a power of the cosmic time, i.e. power-law solutions, or (ii) an exponential of the cosmic time, i.e. a de Sitter (dS) expansion:
 \begin{equation}
a(t)\sim
\begin{cases}
\,t^n\,\,\,\,\,\,\,\mbox{with}\,\,\,\,n\in \mathbb{R}_{\slash 0}&\mbox{(Power-law)}\\
\,e^{\bar{H}t}\,\,\,\,\mbox{with}\,\,\,\,\bar{H}\in \mathbb{R}&\mbox{(de Sitter)}
\end{cases}\,.
\end{equation}
Here,  $\bar{H}$ is the Hubble constant. The present value for the scale factor is assumed to be $a_0 = 1$.
Both the cases are characterized by $\xi\equiv \bar{\xi} = const$. In particular, we have $\bar{\xi}=-1/n$ for power-law solutions, and $\bar{\xi}=0$ for the dS case. Thus, by inversion of Eq. \eqref{eq:weff}, we get the (constant) EoS parameter of the effective fluid to be
 \begin{equation}
 \label{eq:weffcases}
\bweff=
\begin{cases}
-(2/3\,\bar{\xi}+1)\quad&\mbox{(Power-law)}\\
-1\quad&\mbox{(de Sitter)}
\end{cases}\,.
\end{equation}

We recall that the observed present accelerated value for $\bweff$ is $\bweff=-\Omega_\Lambda\approx -0.685$ \cite{Planck:2018vyg}. Instead, an EoS parameter $\bweff=0$ corresponds to an effective matter-dominated era.


\subsection{Pressureless matter $(m)$}
\label{sec:presslessM}

For the illustrative purpose of our study, we decide to assume the simplest case of an energy-momentum tensor dominated by pressureless matter. This type of fluid, which corresponds to an EoS parameter $w_{(m)}=0$, represents baryonic matter and cold dark matter.

In this scenario, let us construct the vector $x = \big(\xi ,\Omega_{(m)}\big)\in\mathbb{R}^2$. It collects the two dynamical variables of our system \eqref{eq:summ}. In particular, using perturbation theory, we can establish up to the first-order the following decomposition
\begin{equation}
\label{eq:pertX}
x\sim \bar{x}+\delta x\,,
\end{equation}
where
\begin{equation}
\bar{x}\equiv \!^\top\!\big(\bar{\xi},\bar{\Omega}_{(m)}\big)
\qquad 
\delta x \equiv \!^\top\!\big(\delta\xi ,\delta\Omega_{(m)}\big)\,,
\end{equation}
with $|\delta\xi|\ll1$ and $|\delta\Omega_{(m)}|\ll1$ small perturbations.
Clearly, this  also shows that any arbitrary function $g(x)$ can be expanded as $ g \sim \bar{g}+\overline{\nabla\!_{x}g}\cdot\delta x$, where the dot denotes cross product.
For the sake of convenience of notation, we will assume hereafter that the crossed-out quantities mean evaluation on $\bar{x}$. Hence, $\overline{\nabla\!_{x}g}=\nabla\!_{x}g|_{\bar{x}}$.

Let us now turn to the modified Friedmann equations with pressureless matter. As a first step, the expansion of equations \eqref{eq:summ1} and \eqref{eq:summ2} with the ansatz \eqref{eq:pertX} 
 leads (i)  to the background (zeroth-order) system 
\begin{subequations}\label{eq:bkpressureless}
\begin{empheq}[left=\empheqlbrace]{align}
	\label{eq:bkpressureless1}
	\, \bar{\Omega}_{(m)} &= \bar{\psi}_0\,,\\	
	\label{eq:bkpressureless2}
	\,\bar{\Omega}'_{(m)} &=\bar{\chi}_{(m)} \bar{\Omega}_{(m)} \,,
\end{empheq}
\end{subequations}
and (ii) to the  first-order matrix equation
\begin{equation}
\label{eq:frst}
\delta x^\prime = \bar{\Pi}\, \delta x\,,
\end{equation}
where
\begin{equation}
\label{eq:piM}
\bar{\Pi}\equiv
\left(
     \begin{array}{cc}
     -\bar{\psi}_1^{-1}\overline{\partial_{\xi }\psi_0} &  \bar{\psi}_1^{-1}\\[0.1cm]
     \bar{\Omega}_{(m)}\overline{\partial_{\xi }\chi_{(m)}} & \bar{\chi}_{(m)}
     \end{array}
     \right)\,.
\end{equation}

In particular, the zero-order system \eqref{eq:bkpressureless}  represents the underlying Friedmann equations that govern the evolution of the universe. In contrast, the first-order equation \eqref{eq:frst} will be useful for performing the linear stability analysis of the background solutions that we will study


\subsubsection{Background equations}

Let us discuss about solutions of the background system \eqref{eq:bkpressureless}. Since $\bar{\psi}_0$ is a constant function, then Eq. \eqref{eq:summPressureless1} dictates that $\bar{\Omega}_{(m)}$ is a constant function too. Therefore, we can rewrite the above system as
\begin{subequations}\label{eq:system}
\begin{empheq}[left=\empheqlbrace]{align}
	\label{eq:summPressureless1}
	\, &\bar{\Omega}_{(m)} = \bar{\psi}_0\,,\\	
	\label{eq:summPressureless2}
	\,&\bar{\chi}_{(m)} \bar{\Omega}_{(m)} =0\,.
\end{empheq}
\end{subequations}

We begin our discussion with equation \eqref{eq:summPressureless2}, identifying two main classes of background solutions to study.

The first class {\small (I)} is characterized by the condition $\bar{\chi}_{(m)}=2\bar{\xi}-3=0$, leading to the solution \eqref{eq:weffcases}
\begin{equation}
\label{eq:DMscenario}
\bar x_{\mbox{\tiny I}}^{\mbox{\tiny DM}}=\!^\top\!\left(-\frac{3}{2},1+\frac{\alpha}{4+\beta}\right) \qquad \bweff^{\mbox{\tiny I,DM}}=0\,,
\end{equation}
with $\beta \ne-4$. Interestingly, this condition is already fulfilled as our system is devoid of singularities by condition \eqref{eq:singfree}. This solution corresponds to an exact matter-dominated era in which $A$ acts as a form of dark matter (DM) in addition to a pressureless matter (dust) contribution, whose density parameter is $\bar{\Omega}_{(m)}^{\mbox{\tiny I}}=\bar{\psi}_0(-3/2)$. Of course, one can also easily see that if $\beta=0$, then the well-know result of Amendola et \emph{al.} obtained in Ref. \cite{AMENDOLA2020135923} for $\mathcal{L}_g=R-\alpha A^{-1}$ is recovered.

Alternatively, the second case {\small (II)} that solves Eq. \eqref{eq:system} is characterized by the condition $\bar{\Omega}_{(m)}^{\mbox{\tiny II}}=0$, i.e a null pressureless matter contribution to universe's energy density. In particular, such case reduces to the  modified FLRW solution
\begin{equation}
\bar x_{\mbox{\tiny II}}=\!^\top\!\left(\bar{\xi}_{\mbox{\tiny II}},0\right) \qquad \bweff^{\mbox{\tiny II}}=-1-\frac{2}{3}\bar{\xi}_{\mbox{\tiny II}}\,,
\end{equation}
where $\bar{\xi}_{\mbox{\tiny II}}$ are the real solutions of the algebraic 4-degree equation $\bar\psi_0(\bar{\xi}_{\mbox{\tiny II}})=0$, whose explicit expression is
\begin{align}
\label{eq:pol4complex}
0&=\big[9\alpha(\beta + 16) + \beta(9\beta + 320) + 2560\big]\nonumber\\
&+24\big[(\beta ( \alpha +  \beta) + 10 \alpha)\big]\bar\xi_{\mbox{\tiny II}}\nonumber\\
&+2\big[\beta (11(\alpha + \beta) + 364)+ 2(37\alpha + 1054)\big]\bar\xi^2_{\mbox{\tiny II}}\\
&+8\big[\alpha ( \beta + 5) + \beta (\beta + 36) + 320\big]\bar\xi^3_{\mbox{\tiny II}}\nonumber\\
&+\big[\alpha(\beta + 4) + (\beta + 20)^2\big]\bar\xi^4_{\mbox{\tiny II}}\,.\nonumber
\end{align}

Of course, directly addressing the $\bar\xi_{\mbox{\tiny II}}$-solutions of Eq. \eqref{eq:pol4complex} is not so simple.
Therefore, instead of discussing the (very long) explicit solutions of such equation on varying $\alpha,\beta$ parameters, we proceed  (i) by first selecting  the $\alpha,\beta$ values that impose two $\bar\xi_{\mbox{\tiny II}}$-solutions of Eq. \eqref{eq:pol4complex} to be the cosmologies we wish to reproduce, and (ii) then  directly solving the residual algebraic 2nd-degree equation.

First of all, to lower the degree of the equation Eq. \eqref{eq:pol4complex},  we decide to force one solution to be a dS state. On the other hand, let's remember that the second no-go theorem for Ricci-inverse gravity considers the dS case, which then emerges as an intriguing solution. Moreover, as we shall see in Subsection \ref{sec:lsa}, this choice will turn out to be particularly apt, being such cosmology a stable attractor of our model.
 
 So, following the  prescription outlined before, one can straightforwardly decide that a first solution could be
\begin{equation}
\label{eq:sol1}
\bar x^{\mbox{\tiny dS}}_{\mbox{\tiny II}}=\!^\top\!\left(0,0\right) \qquad \bweff^{\mbox{\tiny II,dS}}=-1\,.
\end{equation}

Moving to a second cosmological solution, we can't help but think about the observed present  value $\bweff=-\Omega_\Lambda\approx -0.685$ \cite{Planck:2018vyg}. Hence, by using Eq. \eqref{eq:weffcases}, such scenario corresponds to $\bar{\xi}^{\mbox{\tiny DE}}_{\mbox{\tiny II}}\approx-0.473$, which in terms of $\bar{x}$ reduces to the following second solution
\begin{equation}
\label{eq:sol2}
 \bar x^{\mbox{\tiny DE}}_{\mbox{\tiny II}}=\!^\top\!\left(-0.473,0\right)\qquad \bweff^{\mbox{\tiny II,DE}}=-0.685\,.
\end{equation}

Using solutions \eqref{eq:sol1},\eqref{eq:sol2}  and subsequently computing them into the polynomial equation \eqref{eq:pol4complex}, the following values for $\alpha$ and $\beta$ can be obtained
\begin{equation}
\label{eq:albeta}
\alpha \approx-0.028\qquad \beta \approx -15.972\,.
\end{equation}

Interestingly, the value found for $\alpha$ is very close to zero. As far as our $\mathcal{L}_g$ model is concerned,  this fact results in a very week coupling of the anti-curvature contribute $(\beta+AR)^{-1}$ to the gravitational parameter $G$. However, although small, this value contributes significantly to the reproduction of the present accelerated expansion of universe. Moreover, the value of $\beta$  - to our delight  - satisfies the condition \eqref{eq:singfree} that makes our model singularity-free with respect to the cosmological solutions $\bar x^{\mbox{\tiny dS}}_{\mbox{\tiny II}}$ and $\bar x^{\mbox{\tiny DE}}_{\mbox{\tiny II}}$.

With these values in mind, Eq. \eqref{eq:pol4complex} becomes
\begin{equation}
\label{eq:revII}
\xi\left(\xi-\bar{\xi}^{\mbox{\tiny DE}}_{\mbox{\tiny II}}\right)\left(\xi^2+\varepsilon_1\xi+\varepsilon_0\right)=0\,,
\end{equation}
with $\varepsilon_1\approx-0.271$ and $\varepsilon_1\approx0.514$. The solutions of the residual 2nd-degree polynomial equation appearing in Eq. \eqref{eq:revII}  are $\bar{\xi}^{\pm}_{\mbox{\tiny II}}\approx 0.135 \pm0.704 j$. It is clear that such solutions are complex numbers, so they are not physical. We recall that $j$ is the imaginary unit.

We can now study qualitatively our solutions by determining their stability.
%

\subsubsection{Linear stability analysis}
\label{sec:lsa}

In this subsection we will use the linear stability theory to investigate the stability of our cosmological solutions $\bar x^{\mbox{\tiny DM}}_{\mbox{\tiny I}}$, $\bar x^{\mbox{\tiny dS}}_{\mbox{\tiny II}}$ and $\bar x^{\mbox{\tiny DE}}_{\mbox{\tiny II}}$, which may been seen as critical points for the dynamical system \eqref{eq:summ}. 
 This method studies the eigenvalues  $\bar\lambda$ of the matrix $\bar{\Pi}$ defined in \eqref{eq:piM} and, in particular, analyzes the sign of their real part.  On general grounds, if $\mbox{Re}\bar \lambda<0$, $\forall \bar \lambda$, then $\bar x$ is considered  a stable attractor solution of the dynamical system.

By definition, the eigenvalues associated to $\bar{\Pi}$ can be easily determined by finding the roots of the characteristic polynomial
\begin{align}
p_{\bar{\Pi}}(\bar{x},\bar\lambda)&\equiv\det\left(\bar{\Pi}-\bar\lambda\mathbb{I}_2\right)\\
&=\left(\bar{\psi}_1^{-1}\overline{\partial_{\xi }\psi_0}+\bar\lambda\right)\!\left(\bar{\chi}_{(m)}-\bar\lambda\right)+\bar{\psi}_1^{-1}\bar{\Omega}_{(m)}\overline{\partial_{\xi }\chi_{(m)}}\,.\nonumber
\end{align}
As the notation suggests, the characteristic polynomial takes on a different expression depending on the solution $\bar x$ considered. So, let us now move onto the specific cases.

We begin from the DM scenario \eqref{eq:DMscenario}. Searching for the eigenvalues solving
 $p_\Pi(\bar x_{\mbox{\tiny I}}^{\mbox{\tiny DM}},\bar \lambda_{\mbox{\tiny I}}^{\mbox{\tiny DM}})=0$, we get 
\begin{equation}
\bar\lambda_{\mbox{\tiny I},\pm}^{\mbox{\tiny DM}}=\frac{3}{4}
\pm\frac{1}{8\alpha}\sqrt{-6\alpha\big[\alpha(\beta - 2) + (\beta+4)^2\big]}\,.
\end{equation}
After somewhat tedious algebra, the coexistence of the stability condition $\bar\lambda_{\mbox{\tiny I},\pm}^{\mbox{\tiny DM}}<0$ with the singularity-free condition \eqref{eq:singfree} leads to the following constraint
\begin{equation}
\label{eq:DMconstralbet}
-\frac{1}{2}\left(\alpha+8 + \sqrt{\alpha(\alpha + 24)}\right) < \beta < -(\alpha + 4)\,,
\end{equation}
with $0 < \alpha < 8$. In this situation, one can be easily prove that our parameters \eqref{eq:albeta} do not satisfy the constraint \eqref{eq:DMconstralbet}.
This demonstrates that $\bar x^{\mbox{\tiny DM}}_{\mbox{\tiny I}}$ is not a stable solution for our model.

In the same way, for the case {\small (II)} we find that the roots of $p_\Pi(\bar x_{\mbox{\tiny II}}^{\mbox{\tiny DE,dS}},\bar \lambda_{\mbox{\tiny II}}^{\mbox{\tiny DE,dS}})=0$ are
\begin{equation}
\bar\lambda_{\mbox{\tiny II},1}^{\mbox{\tiny DE,dS}}=-\big(1+3\bar\xi^{\mbox{\tiny DE,dS}}_{\mbox{\tiny II}}\big)
\quad
\bar\lambda_{\mbox{\tiny II},2}^{\mbox{\tiny DE,dS}}=-\big(2+3\bar\xi^{\mbox{\tiny DE,dS}}_{\mbox{\tiny II}}\big)\,.
\end{equation}

Again, after somewhat tedious algebra, we obtain that the overlap between the stability domain $\bar\lambda_{\mbox{\tiny II},1}^{\mbox{\tiny DE,dS}}<0$, $\bar\lambda_{\mbox{\tiny II},2}^{\mbox{\tiny DE,dS}}<0$ and the singularity-free condition \eqref{eq:singfree} is:
\begin{equation}
\label{eq:condtri}
\bar{\xi}^{\mbox{\tiny DE,dS}}_{\mbox{\tiny II}}>-\frac{3}{2}\,,
\end{equation}
with $-16<\beta<-4$. In this situation, one can be easily see that 
our solutions are fully consistent with the above stability constraint \eqref{eq:condtri}.
This demonstrates that $\bar x^{\mbox{\tiny dS}}_{\mbox{\tiny II}}$ and $\bar x^{\mbox{\tiny DE}}_{\mbox{\tiny II}}$ are stable attractors of our model.

For a general overview, our findings are summarized in Table I.

\begin{figure}[t]
\includegraphics[width=9.5cm]{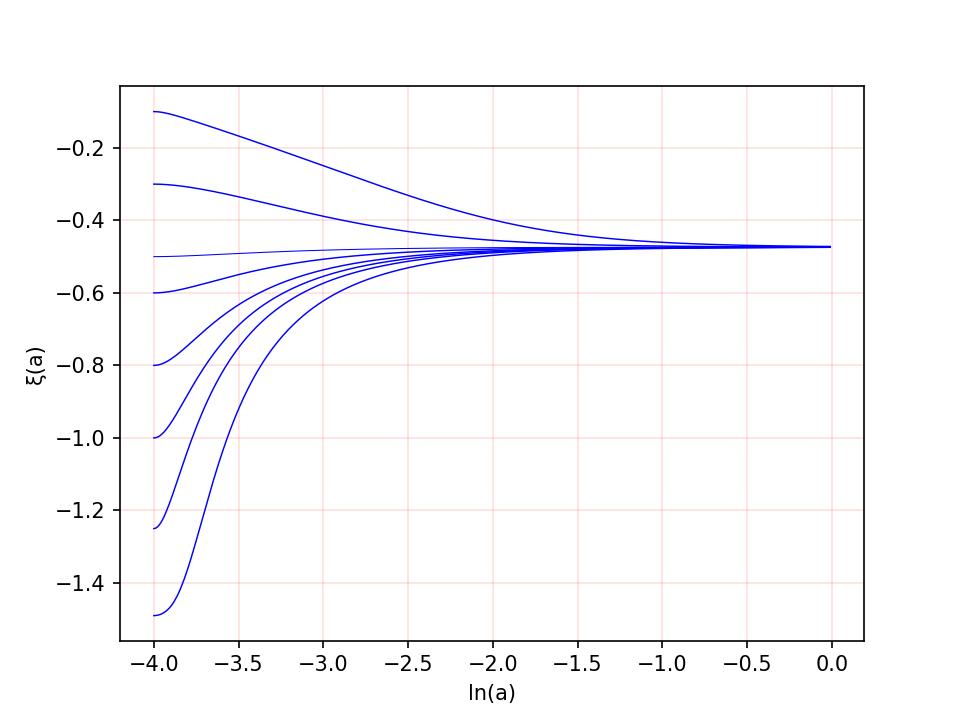}
\centering
\caption{Numerical solutions $\xi(a)$ for Eq. \eqref{eq:summ2} in the pressureless matter case. Here, $\alpha=-0.028$ and $\beta=-15.972$ satisfy both the singularity-free condition \eqref{eq:singfree} and the stability conditions. It is evident that (i) the observed accelerated expansion $\bar{\xi}=-0.472$ is a stable attractor, as well as (ii) the decelerated phase $\bar{\xi}=-1.5$ is an unstable solution. As expected, the cosmic trajectories smoothly join these two epochs without falling into no-go singularities.}
\label{img:aaa}
\end{figure}

Graphical evidence that our model works well is shown in Fig. \ref{img:aaa}, where we solved numerically the evolution equation \eqref{eq:summ2} for $\xi(a)$. In particular, we observe very good agreement with the required behavior of a cosmic expansion moving from a decelerated phase (around $\bweff=0$) to an accelerated phase ($\bweff=-0.685$) without falling into no-go singularities. Moreover, it is evident that the cosmology describing the observed expansion of our universe is a stable attractor.

A further observation can be made.
Since $\bar x^{\mbox{\tiny DM}}_{\mbox{\tiny I}}$ is not a stable solution, the event that may have initiated cosmic evolution from this era may have been a small perturbation of the coupling constant $\alpha$.
To prove this, suppose we start from GR theory (recovered from our model with $\alpha_{\mbox{\tiny GR}}=0$), where the matter-dominated era $\bweff=0$ is a stable cosmological solution. On the basis of the parameterization of our model, suppose we now introduce a small perturbation on $alpha_{\mbox{\tiny GR}}$ such that $\alpha_{\mbox{\tiny GR}}\rightarrow \alpha_{\mbox{\tiny GR}}+\delta\alpha$ with $\delta\alpha\approx-0.028$.
This fact, as seen in the previous discussion, makes the $\bweff=0$ epoch unstable. Therefore, as a result of such a perturbation, our Ricci-inverse \eqref{eq:theory} is established as a new MG theory in which the anti-curvature $A$, acting as a form of DE, smoothly join that phase with the current expansion. Of course, this involves a change in the coupling constant $\kappa$, i.e., the gravitational parameter $G$, of the type $\kappa=\kappa_{GR}+\delta\kappa(\alpha)$.

As a final remark, we point out that if in our model we had considered a different observed value, however close to $-0.7$, for $\bweff$, then the deductions obtainable, apart from the specific values for $\alpha$ and $\beta$, would be the same as in the case discussed in this paper. We also leave to future work any discussion of the existence of
existence of ghosts or other types of instabilities related to our model.
\begin{table}[htp]
\label{tableres}
\begin{center}
\begin{tabular}{|c|l|l|l|c|}
\hline
DM&$\bweff=0$&$\bar{\xi}=-1.5$&$\bar{\Omega}_{(m)}=1.002$&Unstable\\
\hline
dS&$\bweff=-1$&$\bar{\xi}=0$&$\bar{\Omega}_{(m)}=0$&Stable\\
\hline
DE&$\bweff=-0.685$&$\bar{\xi}=-0.472$&$\bar{\Omega}_{(m)}=0$&Stable\\
\hline
\end{tabular}
\caption{Constant solutions for Eq. \eqref{eq:bkpressureless} with $\alpha \approx-0.028$ and $\beta\approx -15.972$. Different epochs of the universe dominated by effective fluids of dark matter (DM), dark energy (DE) and de Sitter (dS) are considered.}
\end{center}
\end{table}


\section{Conclusions}
\label{sec:concl}

In the present work, a novel MG theory called Ricci-inverse gravity was investigated. This theory is based on the introduction of a new, purely geometric object called the anti-curvature tensor, defined as the inverse of the well-known Ricci tensor. 
From this foundation, it is possible to modify the EH action without introducing new fields or requiring a fine-tuning of the coupling constants. 

Although many studies emphasize the increasing relevance and applicability of Ricci-inverse gravity in diverse scenarios, some analyses have fed the idea that Ricci-inverse gravity is not a promising theory to describe the current accelerated expansion of the universe. Foremost among them, later translated into terms of a blocking no-go theorem, is the presence of singularities along the path of cosmic trajectories connecting a decelerated (matter-dominated) epoch to the current accelerated expansion of the universe. 
By requiring the Lagrangian density of such theories to satisfy a new property that we have called regularization via complexification of singularities ($\mathcal{P}2$), we have shown how it is possible to evade the effects of the aforementioned no-go theorem and reconsider cosmologies based on anti-curvature. 

In addition to providing a formal introduction to $\mathcal{P}2$, in this paper we have designed and analyzed for the first time a new scale-free Ricci-inverse gravity model that, by implementing $\mathcal{P}2$, (i) is not rent by no-go singularities and (ii) brings out the current observed expansion as a stable attractor solution. In particular, after obtaining the modified Friedmann equations, we approached the study of their solutions from both an analytical and numerical point of view, giving evidence that they are regular up to the present observed values. 

We trust that, with this analysis, we can once again reopen the discussion aimed at reconsidering gravity theories based on anti-curvature as good dark energy models.

We leave for future investigation the analysis of ghost or possible other instabilities.


\acknowledgments

The author would like to thank Dr. E. Foltran for computational support during the calculations. Some algebraic computations in this article were performed using the Cadabra software \cite{Peeters:2006kp}.


\bibliography{eqmg}

\end{document}